\documentclass[twocolumn,showpacs,preprintnumbers,amsmath,amssymb,floatfix,superscriptaddress]{revtex4-1}
\usepackage{graphicx}
\usepackage{amsmath,amsfonts,amssymb}
\usepackage{graphicx}
\usepackage{color}

\usepackage[normalem]{ulem}
\usepackage{todonotes}
\usepackage[bookmarks,
            breaklinks,
            pdfstartview=FitH, % fit width when the document is opened
            pdftitle={},
            pdfauthor={},
            pdfkeywords={}
            ]{hyperref}
\usepackage{ulem}

\def\beq{\begin{equation}}
\def\eeq{\end{equation}}

\newcommand{\rr}[0]{\mathbf r}
\newcommand{\RR}[0]{\mathbf R}
\newcommand{\pp}[0]{\mathbf p}

\newcommand{\bR}[0]{\mathbf R}

\newcommand{\bra}[1]{\langle #1|}
\newcommand{\ket}[1]{|#1\rangle}

\newcommand{\bs}[0]{\hat{\mathbf S}}

\newcommand{\affA}{Institute of Physics, University of Amsterdam, 1098 XH Amsterdam, The Netherlands}
\newcommand{\affZ}{Eindhoven University of Technology, Post Office Box 513, 5600 MB Eindhoven, The Netherlands}

\newcommand{\appropto}{\mathrel{\vcenter{
  \offinterlineskip\halign{\hfil$##$\cr
    \propto\cr\noalign{\kern2pt}\sim\cr\noalign{\kern-2pt}}}}}

\begin{document}

%\title{A quantum particle in a periodic potential of even- and odd-wave interactions}
\title{Trapped ions in Rydberg-dressed atomic gases}

\author{T.~Secker}\affiliation{\affA}\affiliation{\affZ}
\author{N.~Ewald}\affiliation{\affA}
\author{J.~Joger}\affiliation{\affA}
\author{H.~F\"urst}\affiliation{\affA}
\author{T.~Feldker}\affiliation{\affA}
\author{R.~Gerritsma}\affiliation{\affA}

\begin{abstract}
We theoretically study trapped ions that are immersed in an ultracold gas of Rydberg-dressed atoms. By off-resonant coupling on a dipole-forbidden transition, the adiabatic atom-ion potential can be made repulsive. We study the energy exchange between the atoms and a single trapped ion and find that Langevin collisions are inhibited in the ultracold regime for these repulsive interactions. Therefore, the proposed system avoids recently observed ion heating in hybrid atom-ion systems caused by coupling to the ion's radio frequency trapping field and retains ultracold temperatures even in the presence of excess micromotion. %The scheme also prevents reactive or spin-changing inelastic collisions and can be used for a wide range of atom-ion mass ratios. The results may be of use when studying hybrid atom-ion systems in the ultracold quantum regime.
\end{abstract}

\date{\today}

\maketitle

{\it Introduction} -- Recent years have seen significant interest in coupling ultracold atomic and ionic systems~\cite{Grier:2009,Zipkes:2010,Schmid:2010,Zipkes:2010b,Harter:2013} with the purpose of realising quantum simulators~\cite{Bissbort:2013,Negretti:2014}, studying ultracold chemistry~\cite{Rellergert:2011,Ratschbacher:2012} and collisions or employing ultracold gases to sympathetically cool trapped ions~\cite{Krych:2010,Krych:2013,Meir:2016}. It has become clear however, that the time-dependent trapping field of Paul traps limits attainable temperatures in interacting atom-ion systems~\cite{Zipkes:2010,Schmid:2010,Nguyen:2012,Cetina:2012, Krych:2013,Chen:2014,Weckesser:2015,Meir:2016,Rouse:2017}. This effect stems from the fast micromotion of ions trapped in radio frequency traps which may add significant energy to the system when short-range (Langevin) collisions with atoms occur.

Here, we theoretically study ionic impurities immersed in a cloud of atoms that are coupled to Rydberg states~\cite{Saffman:2009,Henkel:2010,Pupillo:2010,Honer:2010,Balewski:2014,Hofmann:2014}. By selecting a Rydberg state that is coupled to the ground state via a dipole-forbidden transition, we can change the atom-ion interaction such that it becomes repulsive for ultracold atoms. In this situation, the atoms cannot get close enough to the ion for the trap drive to add energy to the system and no excess heating takes place. Instead, the system slowly thermalizes at a rate that is compatible to the rate expected for an ion trapped in a time-independent trap. We also show that in contrast to ground state (attractive) atom-ion systems, which require the mass of the ion to be larger than that of the atom~\cite{Major:1968,DeVoe:2009,Zipkes:2011,Cetina:2012,Chen:2014,Weckesser:2015,Rouse:2017}, the proposed scheme retains ultracold temperatures for the combinations $^{87}$Rb/$^{171}$Yb$^+$ and $^{87}$Rb/$^{9}$Be$^+$ alike.

\begin{figure}[b!]
\centering
\includegraphics[width=8cm]{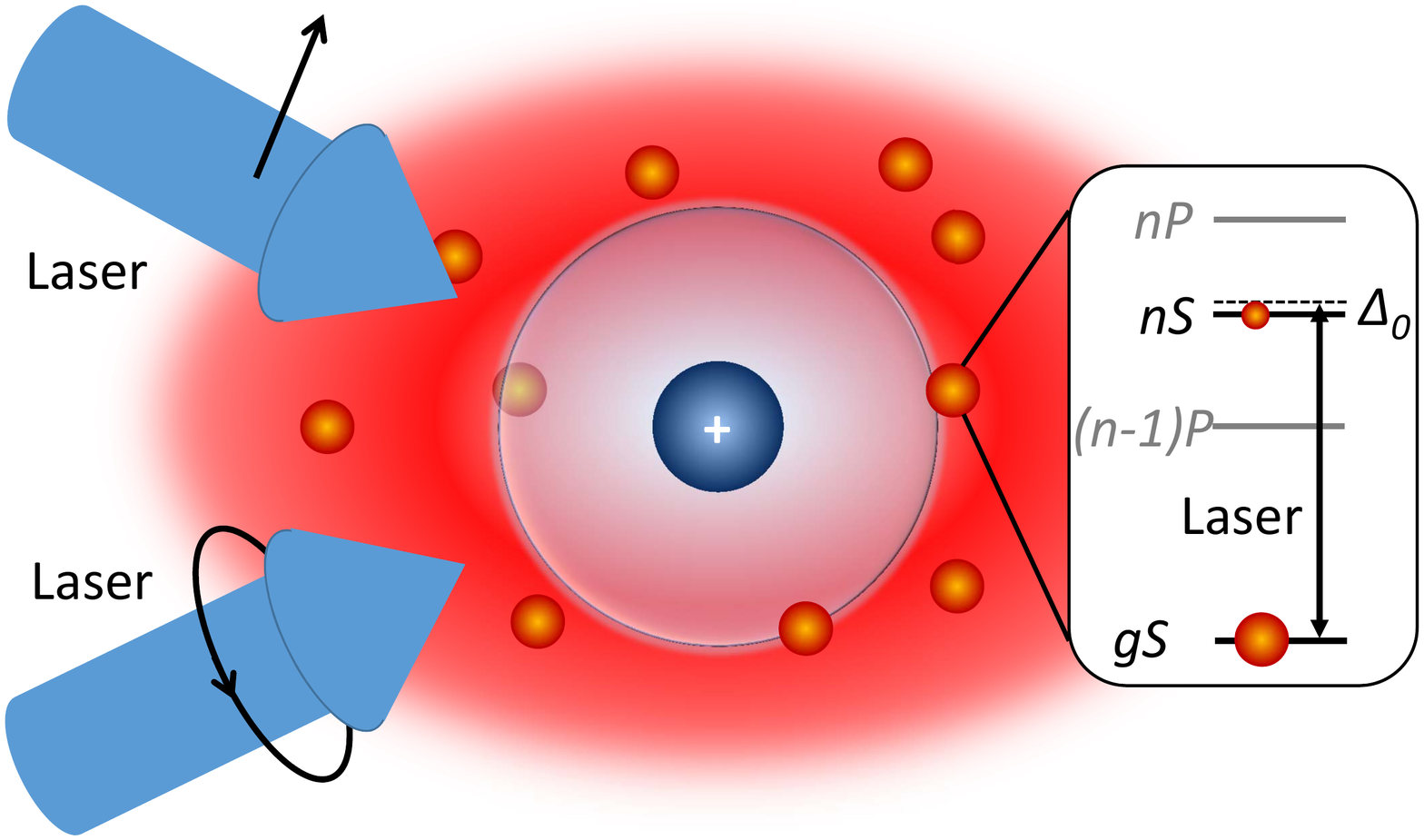}\vspace{-0.5cm}
\caption{A trapped ion (blue sphere) in a cloud of atoms (orange spheres), the atomic ground states $|gS\rangle$ are dressed with the Rydberg state $|nS\rangle$ with $n$ the principal quantum number and $\Delta_0$ the laser detuning. By appropriate engineering of the Rydberg laser field, the adiabatic atom-ion potential can be made repulsive such that the atoms cannot get close enough to undergo Langevin collisions (as depicted by the sphere around the ion). To optically shield the ion in three dimensions, two laser fields are needed with linear and circular polarization as explained in the text. }
\label{fig_blup}
\end{figure}

The paper is organized as follows: First we will describe how the atom-ion interaction can be engineered to be repulsive by coupling to a Rydberg state. Second, we will show that such an atom-ion potential does not lead to heating when the ion is trapped in a Paul trap and is interacting with a cloud of atoms. For this, we will use a classical description of the atom-ion dynamics~\cite{Cetina:2012,Meir:2016}. Finally we discuss possible experimental issues and limitations of our proposed system.

{\it Dressing on a dipole-forbidden transition} -- We are interested in dressing an atom in the ground state $|gS\rangle$ with a Rydberg state $|nS\rangle$ using a single near-resonant laser field, as shown in Fig.~\ref{fig_blup}. When there are no electric fields present, such a transition is prohibited. However, the field of a nearby ion may mix $|nS\rangle$ with Rydberg states that can couple to the $|gS\rangle$ state via a laser field. Within the dipole approximation and using perturbation theory we obtain for the Rydberg wave function:
\begin{equation}\label{psiR}
|\psi (\mathbf{R})\rangle \approx |nS\rangle + e E_{\rm{ion}}(\mathbf{R})\sum_k \frac{\langle kP|z|nS\rangle}{\mathcal{E}_{nS}-\mathcal{E}_{kP}}|kP\rangle,
\end{equation}
\noindent where $\mathcal{E}_j$ denotes the energy of state $|j\rangle$, and $E_{\rm{ion}}(\mathbf{R})$ with $\mathbf{R}=\mathbf{r}_{\rm{i}}-\mathbf{r}_{\rm{a}}$, the electric field generated by an ion located at position $\mathbf{r}_{\rm{i}}$, evaluated at the atomic position $\mathbf{r}_{\rm{a}}$. The relative electron-atom core position projected on the electric field direction is denoted by $z$, where we assume a reference frame in which the local electric field $\mathbf{E}_{\rm{ion}}(\mathbf{R})$ always points in the $z$-direction. In this picture, only states with magnetic quantum number $m_L=0$ are mixed into $|\psi (\mathbf{R})\rangle$ by the field. For further simplification we have neglected fine structure effects for now, which is justified when the laser detuning is much larger than the fine structure splitting of the $|kP\rangle$ Rydberg levels, which will usually be the case. Using expression~(\ref{psiR}) for the Rydberg state, we can estimate the Rabi frequency for one-photon coupling to the ground state $\Omega(\mathbf{R})=e \langle \psi (\mathbf{R})|\mathbf{E}_{\rm{L}}\cdot \mathbf{r}|gS\rangle/\hbar$, where $\mathbf{E}_{\rm{L}}$ denotes the electric field of the laser within the rotating wave approximation and $\mathbf{r}$ the position of the electron in the lab-frame. Combining this equation with eq.~(\ref{psiR}), we get:
\begin{equation}
\Omega(\mathbf{R})\approx \frac{e^2\mathbf{E}_{\rm{L}} \cdot\mathbf{E}_{\rm{ion}}(\mathbf{R})}{\hbar}\sum_k \frac{\langle nS|z|kP\rangle}{\mathcal{E}_{nS}-\mathcal{E}_{kP}}\langle kP|z|gS\rangle.
\end{equation}
Now we can see what happens if we dress the ground state $|gS\rangle$ with $|\psi (R)\rangle$. For large distances $R\rightarrow \infty$, the transition between $|gS\rangle$ and $|\psi (\mathbf{R})\rangle$ will be dipole-forbidden and no dressing occurs. As the distance decreases, we have that the $|nS\rangle$ state gets polarized by the ion and its adiabatic energy shift is given to lowest order by $V_{nS}(R)=-C_4^{nS}/R^4$~\cite{Secker:2016}. We can find the dressed potential by looking at the 2-level Hamiltonian in the $|gS\rangle$, $|\psi (\mathbf{R})\rangle$ subspace after performing the rotating wave approximation:
\begin{equation}
\label{eq:H3level}
H_{\rm{2-level}}=\left( \begin{array}{cc}
-\frac{C_4^{gS}}{R^4} & \hbar\Omega(\mathbf{R}) \\
\hbar\Omega^*(\mathbf{R}) &-\hbar\Delta_0 -\frac{C_4^{nS}}{R^4} \end{array} \right),
\end{equation}
\noindent where we for now neglected any off-resonant couplings and assumed that the laser is blue-detuned by $\Delta_0$ from the $|gS\rangle \rightarrow |nS\rangle$ transition and that this detuning is much smaller than the level splitting between $|nS\rangle$ and all other Rydberg states. In the dispersive limit, where $\hbar|\Omega(\mathbf{R})|\ll \hbar\Delta_0+C_4^{nS}/R^4$ and $C_4^{gS}\ll C_4^{nS}$, diagonalisation results in the following dressed potential:
\begin{equation}\label{eq_dress_simple}
V_{\rm d}(\mathbf{R})\approx\frac{\hbar^2|\Omega(\mathbf{R})|^2}{\hbar\Delta_0+C_4^{nS}/R^4}-\frac{C_4^{gS}}{R^4}.
\end{equation}
In order to proceed, we need to evaluate $|\Omega(\mathbf{R})|^2$, which now depends on the atom-ion separation. We have: $|\Omega(\mathbf{R})|^2=|\beta(\theta,\phi)|^2/R^4$, with:
\begin{equation}\label{eq_beta}
\beta (\theta,\phi )=\frac{e^3E_{\rm{L}}^{\parallel}(\theta,\phi)}{4\pi\epsilon_0\hbar}\sum_k \frac{\langle nS|z|kP\rangle\langle kP|z|gS\rangle}{\mathcal{E}_{nS}-\mathcal{E}_{kP}}.
\end{equation}
\noindent Here, $E_{\rm{L}}^{\parallel}(\theta,\phi)=\mathbf{E}_{\rm{L}} \cdot\mathbf{E}_{\rm{ion}}(\mathbf{R})/|\mathbf{E}_{\rm{ion}}(\mathbf{R})|$ denotes the projection of the laser's electric field onto the electric field of the ion and $\theta$ and $\phi$ denote the azimuthal and polar angle of $\bR$ according to the laboratory reference frame, respectively. Combining equations~(\ref{eq_dress_simple}~and~\ref{eq_beta}), our final result within first order perturbation theory is:
\begin{equation}\label{eq_dress}
V_{\rm d}(\mathbf{R})=\frac{A(\theta,\phi) R_{\rm w}^4}{R^4+R_{\rm w}^4}-\frac{C_4^{gS}}{R^4}.
\end{equation}
The first term in the potential~(\ref{eq_dress}) is repulsive and has a maximum height $A(\theta,\phi)=\frac{\hbar^2|\beta(\theta,\phi)|^2}{C_4^{nS}}$ and width $R_{\rm w}=(C_4^{nS}/(\hbar\Delta_0))^{1/4}$.
The angular dependence of $\beta(\theta,\phi)$ is determined by the laser polarization. For linear polarization along the $z$-direction of the laboratory frame for instance, we have $\mathbf{E}^{z}_{\rm{L}}=(0,0,E_{\rm{L}})$ such that $E_{\rm{L}}^{\parallel}\propto \cos \theta$, whereas circular polarization, $\mathbf{E}^{+}_{\rm L}=E_{\rm L}(1,i,0)/\sqrt{2}$ gives us $E_{\rm{L}}^{\parallel}\propto i\sin\theta\cos\phi +\sin\theta\sin\phi$. The latter results in a donut-shaped repulsive potential in the $x,y$-plane and no interaction along the $z$-direction. In order to make the potential repulsive in all directions we may use two laser fields coupling on the transitions $|gS\rangle\rightarrow|nS\rangle$ and $|gS\rangle\rightarrow|n'S\rangle$ and such that $R_{\rm w}=R'_{\rm w}=(C_4^{n'S}/\hbar\Delta_0')^{1/4}$ with $\Delta_0'$ the bare detuning on the $|gS\rangle\rightarrow|n'S\rangle$ transition. Employing the laser fields $E_{\rm{L}}^z$ and $E_{\rm{L}}^+$ and appropriate tuning of the laser powers, such that the Rabi frequencies are equal, results in a spherically symmetric potential $A(\theta,\phi)=A$, as described in more detail in the Appendix~\ref{Ap_Potential}.

\begin{figure}[t!]
\centering
\includegraphics[width=8cm]{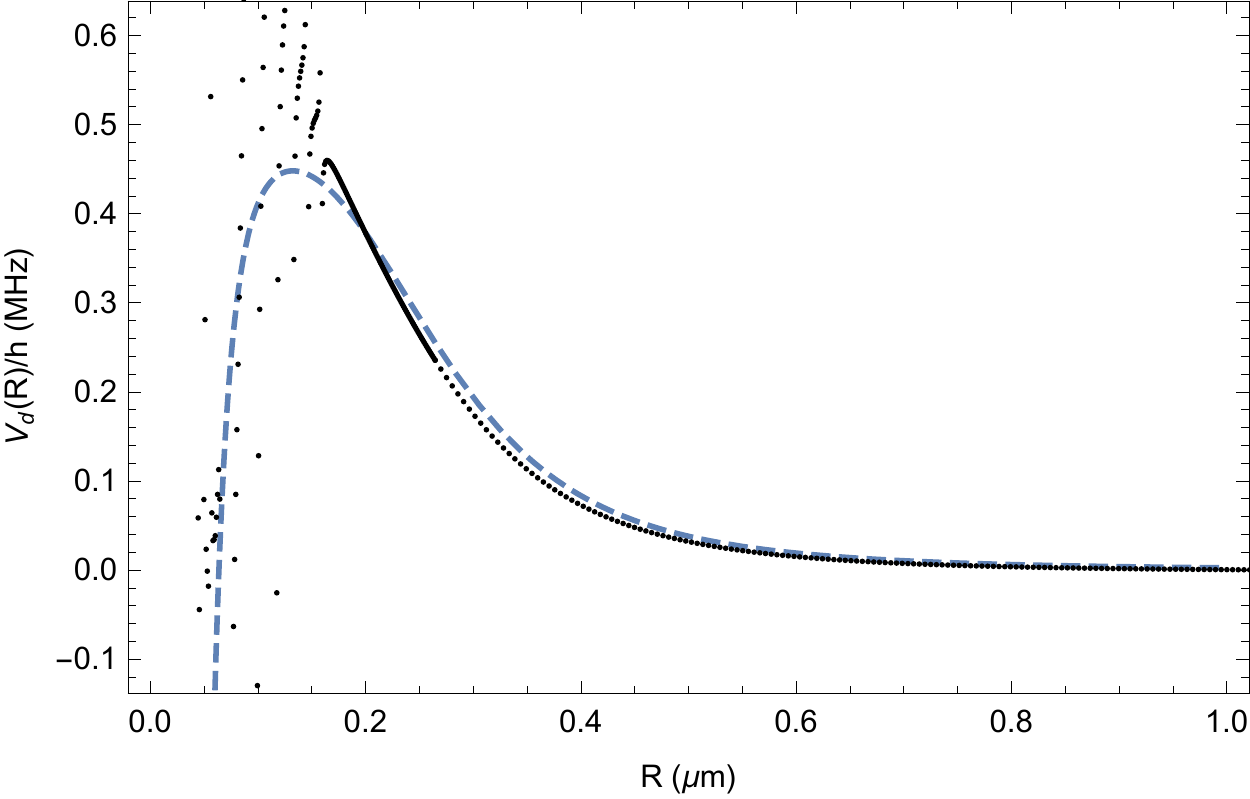}
\caption{The dressed potential $V_{\rm d}(R)$ assuming coupling to the $|19S\rangle$ state of Rb with $\Delta_0=2\pi\times$~1.0~GHz. The dashed line is calculated from the Rydberg wavefunctions taken from Ref.~\cite{Kostelecky:1985} in first order perturbation theory, within the dipole approximation, and neglecting spin-orbit coupling. Here, we assumed the defects of the $L-1/2$ for the $L$ states and set the Rabi frequency on the $|5S\rangle \rightarrow |19P\rangle$ transition to $\Omega_{19P,m_L=0}=2\pi\times$~200~MHz by tuning the laser intensity. The black dots are computed by diagonalization of the Rydberg manifold in the presence of the dressing laser and ion, in the rotating wave approximation, including spin-orbit coupling, up to charge-quadrupole order in the atom-ion interaction terms. The laser intensity was set to the same value as for the approximated potential. For both calculations, we took the Rydberg states $n=10..30$ into account as described in more detail in Appendix~\ref{sec_Diagonalisation}. }
\label{fig_dress}
\end{figure}

An example of a dressed potential calculated within first order perturbation theory is shown in Fig.~\ref{fig_dress}, where we used the approximate analytic wavefunctions of Ref.~\cite{Kostelecky:1985} to evaluate the matrix elements in Eq.~\ref{eq_beta} and $\Delta_0=2\pi\times$~1~GHz such that $R_{\rm w}=$~270~nm. We also show the potential obtained by diagonalization of the Rydberg manifold in the presence of the ion electric field and dressing laser, within the rotating wave approximation, taking spin-orbit coupling and terms up to charge-quadrupole coupling into account. For these calculations, we used the quantum defects found in Ref.~\cite{vanDitzhuijzen:2009} while the Rydberg wavefunctions were obtained by the Numerov method (see Appendix~\ref{sec_Diagonalisation}). This more accurate calculation shows good agreement with our approximated potential for atom-ion separations down to about 160~nm. At close atom-ion proximity, the $|19S\rangle$ state gets strongly coupled to other Rydberg states, causing avoided crossings. To properly compute the dressed potential in this regime, terms beyond the charge-dipole and quadrupole interaction as well as charge exchange effects would have to be taken into account. However, for atom temperatures that are much smaller than the repulsive barrier, $T_{\rm a}\ll A/k_{\rm B}$, short-range collisions between the atoms and ion are suppressed. If we assume a Rabi frequency of $\Omega_{19P,m_L=0}=2\pi\times$~200~MHz on the $|5S\rangle \rightarrow |19P\rangle$ transition in Rb and use that the transition-matrix elements scale as $\langle kP|r|gS\rangle \appropto k^{-3/2}$, we can estimate $A/k_{\rm{B}} \sim 27 \mu$K, which is larger than typical ultracold atom temperatures. We can therefore neglect the inner part of the dressed potential and the second term in equation~(\ref{eq_dress}) for these parameters in the ultracold regime, and for the remaining simulations in this work we replace $V_{\rm d}(\mathbf{R})$ by the spherically symmetric $\tilde{V}_{\rm d}(R)=\frac{A R_{\rm w}^4}{R^4+R_{\rm w}^4}$.

{\it Thermalization in Paul traps} -- To investigate whether the dressed repulsive potential prevents ion heating out of the ultracold regime we have studied classical collisions between an ion and atoms. We assume that we have an ion trapped in a Paul trap generated by the electric fields $\mathbf{E}_{\rm{PT}}(\mathbf{r},t)=\mathbf{E}_{\rm s}(\mathbf{r})+\mathbf{E}_{\rm rf}(\mathbf{r},t)$ with
\begin{eqnarray}
\mathbf{E}_{\rm s}(x,y,z)&=&\frac{m_{\rm{i}}\omega_z^2}{e}\left(\alpha_x\frac{x}{2},\alpha_y\frac{y}{2},-z\right)\label{eqS},\\
\mathbf{E}_{\rm rf}(x,y,z,t)&=&\frac{m_{\rm{i}}\Omega_{\rm rf}^2q}{2e}\cos (\Omega_{\rm rf} t)\,\left(x,-y,0\right)\label{eqRF}.
\end{eqnarray}
\noindent Here, $\Omega_{\rm rf}$ is the trap drive frequency and $q$ is the dynamic stability parameter for an ion of mass $m_{\rm i}$ and charge $e$. The motion of the ion in the transverse $x,y$-direction is given by a slow secular motion of frequency $\omega_{x,y}\approx \frac{\Omega_{\rm rf}}{2}\sqrt{a_{\rm s}+q^2/2}$
with $a_{\rm s}=-\alpha_{x,y}2\omega_z^2/\Omega_{\rm rf}^2$ the static stability parameter, and a fast micromotion of frequency $\Omega_{\rm rf}$. The unitless parameters $\alpha_x+\alpha_y=2$ are introduced to lift the degeneracy in the transverse confinement, as is common in ion trap experiments. The ion dynamics in the $z$-direction is purely harmonic with trap frequency $\omega_z$.

To begin with, we assume that the ion has initially no energy and sits at the center of the Paul trap as in references~\cite{Cetina:2012,Meir:2016}. The idea behind this approach is that although each individual system may be prepared in the ultracold regime by e.g.~laser- and evaporative cooling, it is only when the atoms and ions are brought together that excessive heating is observed~\cite{Meir:2016}. In order to simulate the effect of the atomic cloud we calculate classical trajectories of Rb atoms, with the atoms appearing one-by-one on a sphere of radius 2~$\mu$m around the ion and with a velocity that is randomly picked from a Maxwell-Boltzmann distribution of temperature $T_{\rm a}$. As soon as the atom has left the sphere around the ion at the end of the collision, we introduce a new atom to interact with the ion. We assume the atomic bath to be very large such that its temperature does not change as more collisions occur. After each collision, we obtain the ion's secular energy by fitting an approximate analytic solution of the equations of ion motion to the numerically obtained orbit (see Appendix~\ref{Ap_cols} and reference~\cite{Meir:2016}). We define the ion temperature to be $T_{\rm ion}=\bar{\mathcal{E}}/(3k_{\rm B})$ with $\bar{\mathcal{E}}$ the total average secular energy of the ion. Assuming $^{171}$Yb$^+$ for the ion, we set $\Omega_{\rm rf}=2\pi\times$~2~MHz and $\omega_x=2\pi\times$~54~kHz, $\omega_y=2\pi\times$~51~kHz, $\omega_z=2\pi\times$~42~kHz.

\begin{figure}[t!]
\centering
\includegraphics[width=8cm]{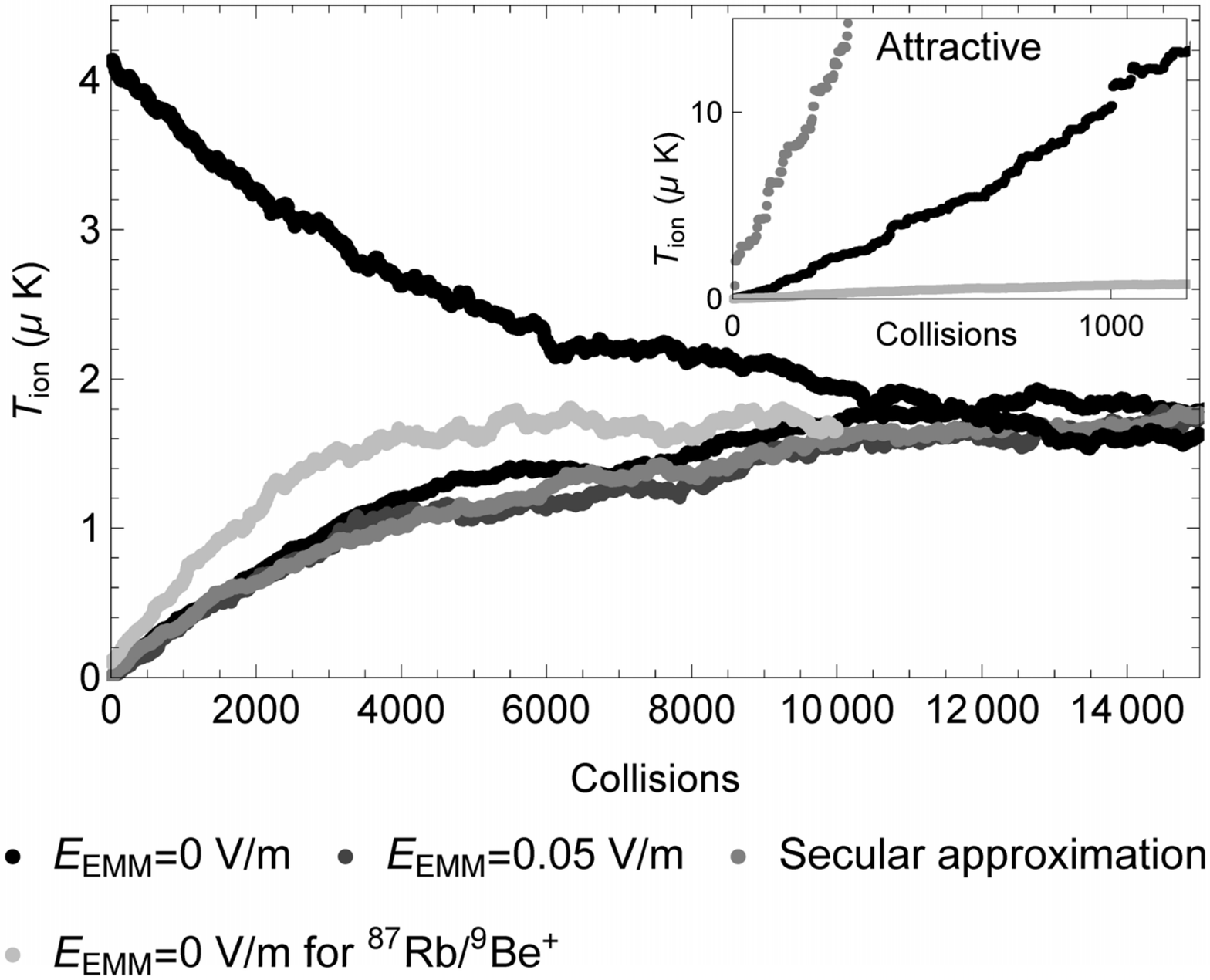}
\caption{  Collision dynamics for an $^{171}$Yb$^+$ ion in a Rydberg-dressed Rb gas with repulsive interactions and $T_{\rm a}=$~2~$\mu$K. The ions heats up from $T_{\rm ion}=0$ to about $1.8$~$\mu$K, corresponding to near-thermalization. Excess micromotion due to a DC field of $E_{\rm EMM}=0.05$~V/m does not affect the thermalization dynamics, which is completely equivalent to that expected for an ion trapped in a time-independent trap with the same secular trap frequencies. For the curve starting at around 4~$\mu$K, we gave the ion an initial velocity of $0.02(1,1,1)$~m/s and we simulate cooling towards the atomic gas temperature. The fastest thermalization curve shows the results for the combination $^{87}$Rb/$^{9}$Be$^+$. The inset shows results for ground state atoms, in which the interaction is attractive. Even without excess micromotion, the ion quickly heats up to temperatures higher than the atomic cloud. This is in stark contrast to the case where we make the secular approximation, where we see that slow thermalization would occur. All results represent the average of 50-100 runs. }
\label{fig_cols}
\end{figure}

We first study the case where the atoms are not dressed to a Rydberg state. Thus, we set the atom-ion interaction to $V_{\rm ai}(R)=-C^{gS}_4/R^4+C_6/R^6$, where the first term denotes the ground state atom-ion induced polarization-charge interaction and the second term accounts for some short-range repulsion, which we set to dominate for $R<10$~nm. The results of averaging 100 of these simulation runs can be seen in the inset of Fig.~\ref{fig_cols}. From the individual trajectories, it is clear that most atoms fly by the ion without causing much effect. One in a few 100 collisions, however, is of the Langevin type causing significant energy exchange. The ion quickly heats up to temperatures much higher than the atomic cloud temperature of 2~$\mu$K. These observations are in line with the findings of references~\cite{Cetina:2012,Meir:2016}. If we simulate more collisions, finite size effects start to limit the accuracy of these calculations, as the ionic oscillation amplitude becomes comparable to the radius of the sphere of atomic starting positions. For comparison, we also show the results obtained when assuming that the ion is in a time-independent trap with the same trap frequencies (the secular approximation). Here, we see that the ion slowly thermalizes with the atomic gas.

Now, we will study how these dynamics change for the Rydberg-dressed atoms for which the atom-ion interaction is repulsive. We replace $V_{\rm ai}(R)$ by $\tilde{V}_{\rm d}(R)$ with $A/k_{\rm B} = 27 \mu$K and $R_{\rm w}=$~270~nm leaving all other parameters in the simulation the same. The results can be seen in Fig.~\ref{fig_cols}: the ion thermalizes slowly with the atomic cloud after about 4$\times 10^3$ ($1/e$) collisions. Starting with an initial ion velocity of $0.02(1,1,1)$~m/s, which corresponds to an initial energy of $\sim$~4~$\mu$K, we can also simulate cooling of the ion towards the temperature of the atomic gas. Up until now, we have assumed ideal experimental conditions in which the minima of $\mathbf{E}_{\rm s}(\mathbf{r})$ and $\mathbf{E}_{\rm rf}(\mathbf{r},t)$ are perfectly overlapped. In practice however, stray electric fields may cause the trap center to not coincide with the radio-frequency null. This causes additional ion motion known as excess micromotion~\cite{Berkeland:1998}. To study its effect, we introduce an additional electric field of $\mathbf{E}_{\rm EMM}=(0.05,0,0)$~V/m, which displaces the potential minimum by $x_{\rm EMM}=$~243~nm corresponding to an average kinetic energy of $\mathcal{E}_{\rm EMM}/k_{\rm B}=m_{\rm i}q^2\Omega^2x_{\rm EMM}^2/(16k_{\rm B})\approx$~100~$\mu$K, and run the simulations. For the repulsive case we find that the dynamics are not affected by the excess micromotion, confirming that the repulsive potential prevents excessive heating of the ion in the Paul trap. Indeed, we repeated the simulation assuming a time-independent potential with the same secular trap frequencies and see no differences beyond statistical fluctuations due to the random atom sampling. Hence, the repulsive potential allows us to make the secular approximation in the hybrid atom-ion system for the parameters used. Note that the assumption that the atom and ion cannot undergo short range collisions still holds in this case as long as the amplitude of oscillation $r_{\rm EMM} =qx_{\rm EMM}/2 \sim$~10~nm~$\ll R_{\rm w}$, even though $\mathcal{E}_{\rm EMM}/k_{\rm B} > A/k_{\rm B}$.

Finally, we study the effect of the atom-ion mass ratio for the repulsive case. For attractive atom-ion interactions it is well-known that stable cold trapping can only be achieved as longs as $m_{\rm a} \lesssim m_{\rm i}$, with $m_{\rm a}$ the  mass of the atom~\cite{Major:1968,DeVoe:2009,Zipkes:2011,Cetina:2012,Chen:2014,Weckesser:2015,Rouse:2017}. To see how this changes for the repulsive case, we repeated the simulations for the combination $^{87}$Rb/$^{9}$Be$^+$, whose mass ratio lies far outside of the range where thermalization can be expected. As can be seen in Fig.~\ref{fig_cols}, this combination also remains within the ultracold regime when the interactions are repulsive, thermalizing after about 2$\times 10^3$ ($1/e$) collisions.

{\it Experimental issues} -- For the calculations in this paper we have neglected the effect of the ionic trapping fields on the dressing of the atom. This is justified since the trapping field at maximal amplitude is much smaller than the field of the ion at the dressing point: $\rm{Max}(E_{\rm PT}(R_{\rm w},t))/E_{\rm ion}(R_{\rm w})\approx $~$10^{-4}$ ($\approx $~$10^{-5}$ for $^9$Be$^+$), such that we can safely neglect these fields. Far away from the center of the trap, however, the maximal amplitude of the trapping fields become similar to $E_{\rm ion}(R_{\rm w})$. For the numbers used in this work, this occurs only at $R\approx$~1~mm from the trap center, such that we can also safely neglect the effect of the dressing on the atomic cloud far away from the ion.

For the simulations, we assumed that only a single atom was interacting with the ion at the same time, which would require a density of $\rho_{\rm a}\leq 1/R_{\rm w}^3\approx 10^{19}$~m$^{-3}$. Assuming an atomic density of $\rho_{\rm a}=10^{18}$~m$^{-3}$ and $T_{\rm a}=$~2~$\mu$K, we have a flux of about 6$\times$10$^5$ atoms/s entering the sphere of $R=2$~$\mu$m, such that each collision in Fig.~\ref{fig_cols} amounts to a few $\mu$s. We estimate an upper bound to the photon scattering rate as $\Gamma\leq f \Gamma_{19S}\rho_{\rm a}R_{\rm w}^3\approx 10$~Hz, for $\rho_{\rm a}=10^{18}$~m$^{-3}$ and with $\Gamma_{19S}$ the decay rate of the Rydberg state and $f$ the probability for an atom to be found in the Rydberg state: $f\propto \hbar^2|\Omega(\mathbf{R})|^2/(\hbar\Delta_0+C_4^{nS}/R^4)^2 \leq 10^{-3}$. Therefore, we expect that photon scattering will not affect the collision dynamics considerably. For increasing ion temperatures, there is a probability that the atom makes it over the repulsive barrier in a collision. To investigate this regime, we simulated 2$\times$10$^5$ collisions between atoms with $T_{\rm a}=2\mu$K and $T_{\rm ion}$ about 100~$\mu$K and found that in 146 instances, the atoms approached the ion to distances below $R_{\rm w}/2$, such that they entered the inner regime of the dressed potential. For $T_{\rm ion}$ around 10.6~$\mu$K, the number of atoms entering the inner part of the potential dropped to zero, as described in more detail in Appendix~\ref{Ap_OverBar}. We conclude that our scheme works best for ions and atoms that are pre-cooled into the ultracold regime before the systems are merged. Finally, we estimate the probability $P_{\rm tunnel}$ for the atom to tunnel through the potential barrier to be $P_{\rm tunnel}\approx {\rm exp}(-\sqrt{2 \mu_{\rm ai} (A-k_{\rm B} T_{\rm a})/\hbar^2}R_{\rm w})\sim 10^{-9}$, with $\mu_{\rm ai}$ the atom-ion reduced mass, such that we can safely neglect this effect.

{\it Conclusions} -- We have shown that trapped ions in Rydberg-dressed gases are stable against micromotion assisted heating when we design the dressed atom-ion potential to be repulsive. To engineer this repulsive interaction, a scheme that employs one-photon dressing on a dipole forbidden transition can be used. While Rydberg states with negative polarizability could also be used to engineer repulsive atom-ion interactions, these are not always available or can be hard to spectrally resolve. In contrast, the scheme proposed in this paper works for any alkali atom and avoids close-by spectator states by employing Rydberg states without orbital angular momentum, which generally have large energy separations from other Rydberg states. The scheme may be employed for a wider range of atom-ion mass ratios than is the case for the attractive atom-ion system. The results may be of interest when studying atom-ion interactions in the quantum regime and may find applications in hybrid quantum computation or simulation approaches~\cite{Doerk:2010,Bissbort:2013,Secker:2016} and the study of polarons~\cite{Casteels:2011} in these systems.

\section*{Acknowledgements}
This work was supported by the EU via the ERC (Starting Grant 337638) and the Netherlands Organization for Scientific Research (NWO, Vidi Grant 680-47-538) (R.G.). We gratefully acknowledge fruitful discussions with Antonio Negretti and Alexander Gl{\"a}tzle.
%-------------------------------
%\bibliography{biblio-RG}
%-------------------------------

\appendix

\onecolumngrid

\section{Spherically symmetric potential}
\label{Ap_Potential}

We can use two dressing lasers to engineer a spherically symmetric potential, coupling on the transitions $|gS\rangle\rightarrow|nS\rangle$ and $|gS\rangle\rightarrow|n'S\rangle$. After performing the rotating wave approximation, we can write the Hamiltonian in the $|gS\rangle$, $|\psi (\mathbf{R})\rangle$, $|\psi' (\mathbf{R})\rangle$ subspace

\begin{equation}
\label{eq:H3level}
H_{\rm{3-level}}=\left( \begin{array}{ccc}
-\frac{C_4^{gS}}{R^4} & \hbar\Omega(\mathbf{R}) & \hbar\Omega'(\mathbf{R})\\
\hbar\Omega^*(\mathbf{R}) &-\hbar\Delta_0 -\frac{C_4^{nS}}{R^4} & 0\\
\hbar\Omega'^*(\mathbf{R})& 0 & -\hbar\Delta'_0 -\frac{C_4^{n'S}}{R^4}\end{array} \right).
\end{equation}

To obtain the adiabatic potential, we diagonalise $H_{\rm{3-level}}$ assuming $\hbar|\Omega(\mathbf{R})|\ll \hbar\Delta_0+C_4^{nS}/R^4$, $C_4^{gS}\ll C_4^{nS}$ and $\hbar|\Omega'(\mathbf{R})|\ll \hbar\Delta'_0+C_4^{n'S}/R^4$, $C_4^{gS}\ll C_4^{n'S}$ so that we can expand to second order in $\hbar|\Omega(\mathbf{R})|$ and $\hbar|\Omega'(\mathbf{R})|$ to find:

\begin{equation}\label{eq_dress3}
V_{3d}(\mathbf{R})\approx\frac{A(\theta,\phi) R_{\rm w}^4}{R^4+R_{\rm w}^4}+\frac{A'(\theta,\phi) {R'}_{\rm w}^4}{R^4+{R'}_{\rm w}^4}-\frac{C_4^{gS}}{R^4}.
\end{equation}

\noindent with $A(\theta,\phi)=\frac{\hbar^2|\beta(\theta,\phi)|^2}{C_4^{nS}}$ and $A'(\theta,\phi)=\frac{\hbar^2|\beta'(\theta,\phi)|^2}{C_4^{n'S}}$. The angular dependence of $\beta$ is expressed as:

\begin{align}\label{eq_beta3}
\beta(\theta,\phi)& =\frac{e^3E_{\rm L}^{\parallel}(\theta,\phi)}{4\pi\epsilon_0\hbar}\sum_k \frac{\langle nS|z|kP\rangle\langle kP|z|gS\rangle}{\mathcal{E}_{nS}-\mathcal{E}_{kP}}\\
\beta'(\theta,\phi)& =\frac{e^3{E'}_{\rm L}^{\parallel}(\theta,\phi)}{4\pi\epsilon_0\hbar}\sum_k \frac{\langle n'S|z|kP\rangle\langle kP|z|gS\rangle}{\mathcal{E}_{n'S}-\mathcal{E}_{kP}}.
\end{align}

\noindent Where $E_{\rm L}^{\parallel}(\theta,\phi)$ denotes the projection of the laser electric field onto the electric field of the ion. Now we set $\mathbf{E}_{\rm L}=(0,0,E_{\rm L})$ such that $\beta(\theta,\phi)= \beta_0\cos \theta$ and $\mathbf{E}'_L={E'}_L(1,i,0)/\sqrt{2}$ such that $\beta'(\theta,\phi)= \beta_0'( i\sin\theta\cos\phi +\sin\theta\sin\phi)$. In this situation,

\begin{equation}\label{eq_dress3}
V_{3\rm{d}}(\mathbf{R})\approx\frac{|\beta_0|^2 \cos^2\theta R_{\rm w}^4}{C_4^{nS}(R^4+R_{\rm w}^4)}+\frac{|\beta'_0|^2 \sin^2\theta {R'}_{\rm w}^4}{C_4^{n'S}(R^4+{R'}_{\rm w}^4)}-\frac{C_4^{gS}}{R^4},
\end{equation}

\noindent which results in a spherically symmetric potential if we set $R_{\rm w}=R'_{\rm w}$ and $|\beta_0|^2/C_4^{nS}=|\beta'_0|^2/C_4^{n'S}$, which can be done by tuning $\Delta_0$, $\Delta'_0$ and the laser intensities.

\section{Determination of ion temperature}\label{Ap_cols}
The explicit time dependence of the electric trapping field obstructs a straight-forward definition of some physical quantities such as energy and temperature. Nonetheless, if the time dependence of the external field is periodic and the underlying equations of motion are linear, we are able to apply Floquet's theory and the problem can be decomposed into a part which oscillates with the same period as the external field and a part, referred to as the secular part, which is time independent~\cite{Leibfried:2003}.
Often useful physical quantities can be constructed from the secular problem and its solution, e.g.~in systems of multiple ions energy and temperature can be defined via the secular solutions of the linearized problem.
For the combined system of atoms and ions we do not have a closed form for the solution at hand, but in the considered case of low densities the system will evolve as the uncoupled one for time periods in which all atoms are spatially well separated from the ion. During those time periods the interaction forces can be neglected and a measurement of the secular energy of the isolated ion system is possible. The ion's secular energy $\cal{E}$ can be determined via its secular frequencies $\omega_i$, $i=x,y,z$, and the amplitude of the orbit's secular component. For a single ion in an oscillating electric field $\mathbf{E}_{\rm PT}(\mathbf{r},t)$, as defined in Eq.~7 of the main text, with appropriately chosen trap parameters the equations of motion can be solved analytically and the solutions take the following approximate form
\begin{align}
x (t) & \approx \left( C_x \cos(\omega_x t) + S_x \sin(\omega_x t) \right) \left( 1 + \frac{q}{2} \cos(\Omega_{\text{rf}} t ) \right) - \frac{\omega_x q}{\Omega_{\text{rf}}} \left( S_x \cos( \omega_x t) - C_x \sin( \omega_x t) \right) \sin( \Omega_{ \text{rf}} t ),\\
y (t) & \approx \left( C_y \cos(\omega_y t) + S_y \sin(\omega_y t) \right) \left( 1 - \frac{q}{2} \cos(\Omega_{\text{rf}} t ) \right) + \frac{\omega_y q}{\Omega_{ \text{rf}}} \left( S_y \cos( \omega_y t) - C_y \sin( \omega_y t) \right) \sin( \Omega_{\text{rf}} t ),\\
z (t) & = \left( C_z \cos(\omega_z t) + S_z \sin(\omega_z t) \right) \, . \phantom{\frac{q}{2}}
\end{align}
Here we neglected correction terms which oscillate faster than the trap drive frequency $\Omega_{\text{rf}}$. Given the ion's position at two instances of time the above system of linear equations can be inverted to obtain the coefficients $C_i$ and $S_i$ of the secular solution's cosine and sine part. As a preliminary step the secular frequencies can be determined numerically by fitting the above analytic approximation to a numerically obtained solution. This increases accuracy as compared to the approximate analytical expression for the $\omega_i$ we gave in the main text. Given all this we can now calculate the secular energy $\mathcal{E}=\sum_{i=x,y,z} m_{\rm i} \omega_{i}^2 (C_i^2+S_i^2) / 2 $ of the isolated ion system.\\
In the case of the repulsive potential, which we analyzed in the main text, the strong similarity of the thermaliztion process for the full and the secular problem indicates that the ion's secular temperature for this type of potentials is indeed a meaningful physical observable, in contrast to the attractive case.

\section{Over barrier collisions}\label{Ap_OverBar}
\begin{figure}[b!]
\centering
%\begin{subfigure}{.49 \textwidth}
%\centering
\includegraphics[width=8.5cm]{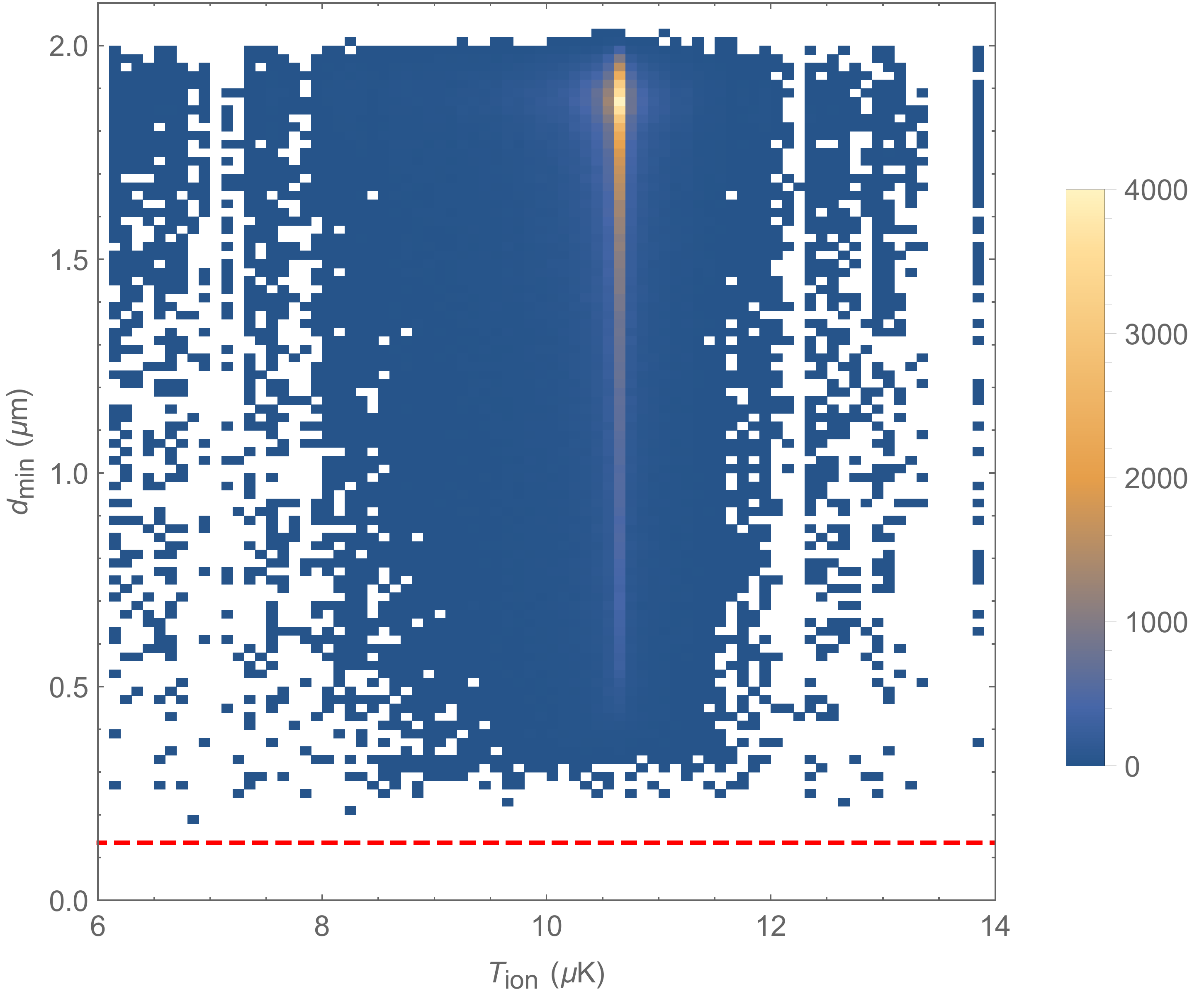}
\caption{Ion temperature $T_{\rm ion}$ vs. minimal atom-ion separation $d_{\rm min}$ for 2~$\times$~10$^5$ collision simulations of Rb and Yb$^+$ as described in the main text with $T_{\rm ion}$ around 10.6~$\mu$K. The dashed line indicates $R_{\rm w}/2$, where the atoms find themselves within the inner part of the dressed potential.}
\label{fig:over_barrier1}
\end{figure}

\begin{figure}
\centering
\includegraphics[width=8.5cm]{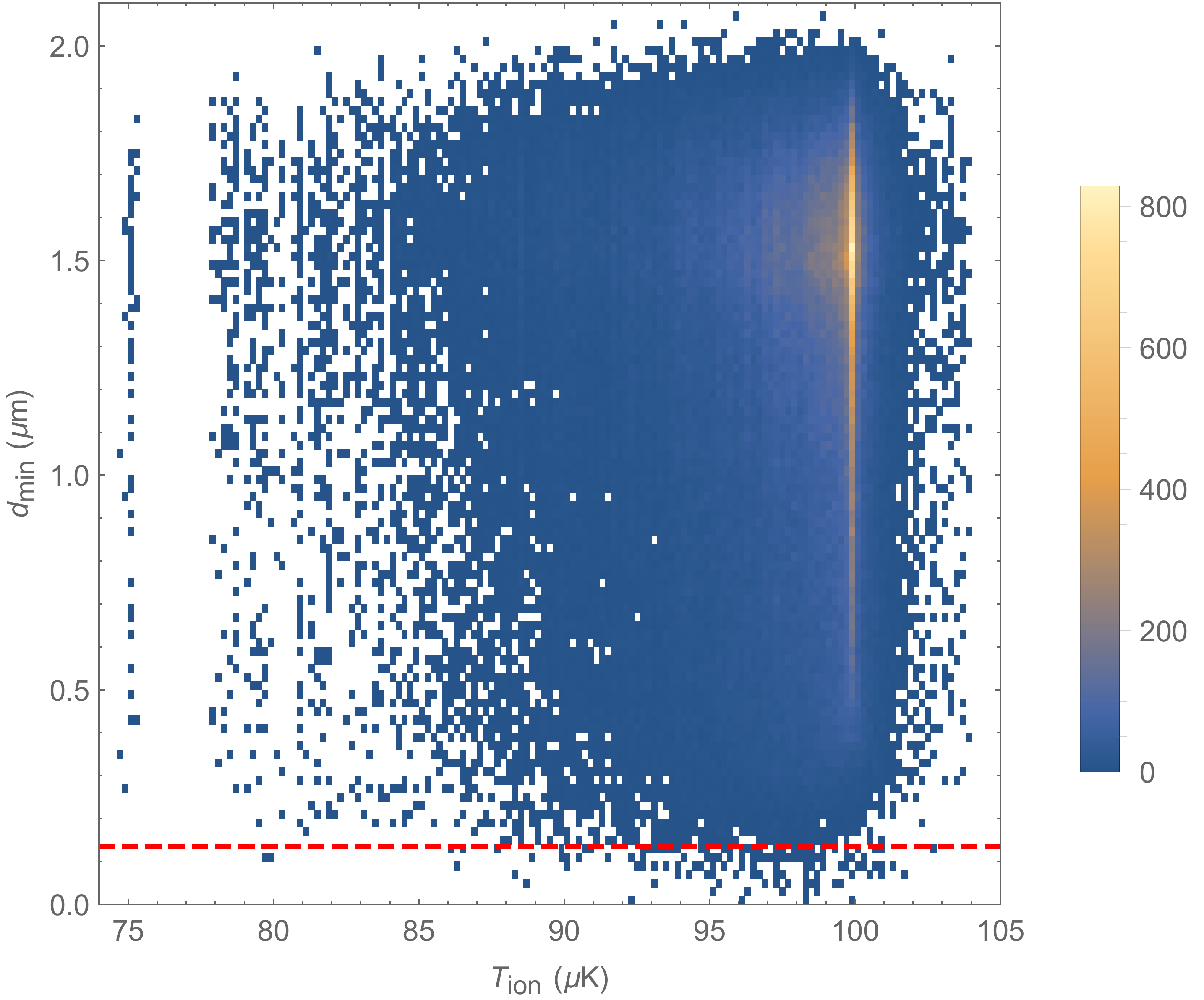}
\caption{Ion temperature $T_{\rm ion}$ vs. minimal atom-ion separation $d_{\rm min}$ for 2~$\times$~10$^5$ collision simulations of Rb and Yb$^+$ as described in the main text with (a) $T_{\rm ion}$ around 100~$\mu$K. The dashed line indicates $R_{\rm w}/2$, where the atoms find themselves within the inner part of the dressed potential.}
\label{fig:over_barrier2}
\end{figure}

In our simulations of the classical collision dynamics there is a non-vanishing probability for the atoms to overcome the repulsive barrier. Since the ion is strongly localized, we estimate this to occur when the amplitude of ion oscillation exceeds $R_{\rm w}$, or $T_{\rm ion}\gtrapprox m_{\rm i}\bar{\omega}^2R_{\rm w}^2/(2k_{\rm B})\approx$~70~$\mu$K for the parameters used in this work, with $\bar{\omega}$ the average trap frequency. To estimate if those events pose a problem we performed collision simulations where we kept record of the minimal separation $d_{\rm min}$ between atom and ion.
More precisely we simulated the interaction of an ion with a bath of $T_a=$2~$\mu$K atoms for 200 collisions. The result of 1000 such simulations for ion temperatures $T_{\rm ion}$ around $10.6$~$\mu$K and $100$~$\mu$K are shown in Fig.~ \ref{fig:over_barrier1} and \ref{fig:over_barrier2}. For $T_{\rm ion}\approx$100~$\mu$K we found that in 146 instances out 2~$\times$~10$^5$ times, the atoms approached the ion to distances below $R_{\rm w}/2$, such that they entered the inner regime of the dressed potential. For $T_{\rm ion}\approx$10.6~$\mu$K, the number of atoms entering the inner part of the dressed potential dropped to zero.

\section{Diagonalization of the Rydberg-ion interaction Hamiltonian}\label{sec_Diagonalisation}
In this section we discuss how we calculated the adiabatic atom-ion potential from a direct-diagonalisational approach. The atom-ion potential can within the adiabatic approximation be obtained from the internal level structure of the atom for different but fixed atom ion separations $\RR$. Each of those $\RR$-dependent energy levels can be viewed as the effective potential between atom and ion, under the assumption, that the internal state follows the corresponding eigenstate adiabatically~\cite{Secker:2016}. We focus on the potential or channel, which correspond to the atom's ground state at infinite separation. The relevant shifts of the ground state will with the parameters we choose and for separations larger than $\sim$~0.2~$\mu$m result from the off-resonant coupling to the Rydberg manifold, as can be seen from the simpler perturbational approach discussed in the main text. Therefore, we neglect the effect of the ion's electric field on the ground state level for now keeping in mind that the resulting potential will be a good approximation just in the range considered above. For the Rydberg manifold on the other hand the presence of the ion leads to relevant corrections, since it introduces strong couplings between the Rydberg states due to the large polarizability $\alpha\propto n^7$ of the Rydberg states, with $n$ the principal quantum number. Therefore, we model the Hamiltonian representing the atom's internal structure by a single ground state $\ket{gS}$ with energy $\mathcal{E}_{gS}$, which we assume to be unaffected by the ion's electric field, coupled by the dressing laser field to a set of Rydberg states, which are affected by the ion field. With those considerations the Hamiltonian representing the atom's internal structure in the presence of ion and dressing laser reads

\beq
\begin{aligned}\label{HBO}
\hat H
%(\RR)
=&\hat{H}_{gS}+\hat{H}_{Ryd}+\hat{H}_{ia}+\hat{H}_{L},\\
\end{aligned}
\eeq

\noindent where $\hat{H}_{gS}$ describes the unperturbed atomic ground state $\ket{gS}$ and is given by:

\beq
\begin{aligned}
\hat{H}_{gS}=&\mathcal{E}_{gS} \ket{gS}\bra{gS}.\\
\end{aligned}
\eeq

\noindent $\hat{H}_{\rm Ryd}$ represents a set of unperturbed atomic Rydberg states in diagonal form, which in quantum defect theory~\cite{vanDitzhuijzen:2009} can be represented by a single Rydberg electron and a singly charged atomic core, which represents the inner electrons and the nucleus:

\beq
\begin{aligned}
\hat{H}_{Ryd}=&\sum_{k} \mathcal{E}_{(k)} \ket{k}\bra{k},\\
\end{aligned}
\eeq

\noindent where $k$ represents the tuple of quantum numbers $(n,l,j,m_j)$ ranging over the common hydrogen fine structure states and $\mathcal{E}_{(k)}$ is the corresponding Rydberg energy level with eigenstate $ \ket{k}$ in the relative Rydberg electron-core variable. We obtained those eigenenergies and states in coordinate representation employing the Numerov method and quantum defects found in literature~\cite{vanDitzhuijzen:2009}.

\noindent $\hat{H}_{\rm ia}$ comprises the interaction terms between Rydberg atom and ion in the quantum defect theory approach. It consists of two Coulomb interaction terms $V_C$, a repulsive one between ion and atomic core and an attractive one between ion and Rydberg electron. In addition, we included a spin-orbit like coupling term $V_{SO}^{e-i}$ representing the coupling of the ion's electric field to the spin of the Rydberg electron~\cite{Secker:2016}.

\beq
\label{eq:atomionpert}
\begin{aligned}
\hat{H}_{ia}=&V_C(\rr_c-\rr_i)-V_C(\rr_e-\rr_i)+V_{SO}^{e-i}\\
=&-\frac{e^2}{4\pi \epsilon_0 |\RR+\frac{m_c}{M}\rr|}
+\frac{e^2}{4\pi \epsilon_0 |\RR-\frac{m_e}{M}\rr|}
-\frac{1}{2 m_e^2 c^2 }\, \bs\cdot\left(\left(\frac{e^2 (\RR+\frac{m_c}{M}\rr)}{4\pi \epsilon_0 |\RR+\frac{m_c}{M}\rr|^3}\right)\times\pp\right)\\
\approx&
\frac{e^2}{4\pi \epsilon_0|\RR|^3}\left(-\rr\cdot\RR+\frac{m_c-m_e}{2M}\rr^2\right)-\frac{3e^2(m_c-m_e)}{8\pi\epsilon_0 M|\RR|^5}(\rr\cdot\RR)^2%\\
%&
-i \frac{e^2\mu}{4\pi\epsilon_0\hbar m_e^2  c^2}\frac{1}{|\RR|^3}\,\bs\cdot(\RR\times[\hat H_{Ryd},\rr]).\\
\end{aligned}
\eeq

\noindent Here $\rr_i$, $\rr_c$ and $\rr_e$ are the ion, core and Rydberg electron position operators in the laboratory frame respectively, $\rr$ and $\pp$ are the relative position and relative momentum operator between core and Rydberg electron, $\bs$ is the spin-1/2 operator of the electron, $e$ denotes the elementary charge, $m_e$ and $m_c$ denote the electron and core mass respectively, $M$ and $\mu$ are the total and reduced mass of the atom respectively and $\epsilon_0$ is the vacuum permittivity.
The approximation has been obtained by taylor expanding the terms and substituting the momentum operator in the last term with $\pp\approx i 2 \frac{\mu}{ \hbar} [\hat H_{Ryd},\rr]$~\cite{Secker:2016}. The approximation includes terms up to quadrupole order in the Coulombic terms taking account for the non-linearity of the potential over the spatial extent of the Rydberg wavefunctions considered.

\noindent The term $\hat{H}_L$ of Eq.~\ref{HBO} describes the dressing laser field in dipole approximation, which couples the Rydberg states off-resonantly to the ground state $\ket{gS}$.
\beq
\begin{aligned}
\hat{H}_L=&\sum_k e^{i\omega_L t}\Omega_{(k,gS)}\ket{gS}\bra{k}+h.c.
\\
\end{aligned}
\eeq

\noindent where $\Omega_{(k,gS)}$ denotes the Rabi frequency of the transition $\ket{k}$ to $\ket{gS}$ and the sum runs over all Rydberg states $(k)=(n,P,j,m_j)$ for which the transition is dipole allowed.

\noindent We proceed in changing to a rotating frame of the ground state with respect to the dressing laser frequency and perform a rotating wave approximation. This affects only $\hat{H}_{gS}$ and $\hat{H}_L$, which transform to
\beq
\begin{aligned}
\hat{H}_{gS}^{'}=&(\mathcal{E}_{gS}+h \omega_L) \ket{gS}\bra{gS}\\
\hat{H}_L^{'}=&\sum_k \Omega_{(k,gS)}\ket{gS}\bra{k}+h.c.
\\
\end{aligned}
\eeq

\noindent To determine the coupling strengths $\Omega_{((n,P,j,m_j),gS)}$ we computed the transition matrixelements with an approximate solution for the ground state wavefunction, which we obtained as in the Rydberg case. We restrict to the case of linear polarization along the direction of the ion electric field and fix the couplings such that they are consistent with the case that neglects the fine structure in the main text, namely $\Omega_{(19P,gS)}=2 \pi\times$~200~MHz. To this end, we rescaled the couplings such that $\Omega_{((19,P,3/2),gS)}=\Omega_{(19P,gS)} A_{FS}/A$ with $A=\bra{P,m_l=0}\cos(\theta) \ket{S,m_l=0}$ and $A_{FS}=\bra{P,j=3/2}\cos(\theta) \ket{S,j=1/2}$ the angular component of the dipole operator coupling states without fine structure and with fine structure states respectively. The couplings obtained in this way were found to be in agreement with the scaling law
$\Omega_{((n,P,j),gS)}\propto n^*(n,P,j)^{-3/2}$ with $n^*(n,l,j)=n-\delta_{nlj}$ and $\delta_{nlj}$ the quantum defects. Diagonalisation for $n=10..30$ and $\omega_L=(\mathcal{E}_{(19,S,1/2)}-\mathcal{E}_{gS})/h+\Delta$ results in the potential shown in Fig.~2 of the main text.


\end{document}